\documentclass[runningheads]{llncs}

\usepackage{amsmath}
\usepackage{amsfonts}
\usepackage{amssymb}
\usepackage{lmodern}
\usepackage{mathtools}

\usepackage{graphicx}
\usepackage{braket}
\usepackage{comment}
\usepackage{xfrac}

\usepackage{booktabs}
\usepackage[caption=false]{subfig}
\usepackage{adjustbox}

\usepackage{qcircuit}

\usepackage{hyperref, xcolor}

\usepackage{pgf}
\usepackage{float}

\newcommand{\cent}[0]{\mbox{\textcent}}
\newcommand{\mymatrix}[2]{\left( \begin{array}{#1} #2 \end{array} \right)}

\newcommand{\mypar}[1]{\left( #1 \right)}

\newcommand{\abs}[1]{\left\lvert#1\right\rvert}

\newcommand{\setBuilder}[2] { 
    \ensuremath{ \{{#1} \mid {#2}\}}
}

\newcommand{\MOD}[1]{\ensuremath{\mathtt{MOD}_{\rm #1}}}
\newcommand{\MODp}[0]{\MOD{p}}

\newcommand{\bigO}{\ensuremath{\mathcal{O}}}

\newcommand{\scl}[1]{\scalebox{0.9}{#1}}

\begin{document}
\title{Implementing Quantum Finite Automata Algorithms on Noisy Devices}
%
%
\author{Utku Birkan\inst{1,2} \and
Özlem Salehi\inst{3,7} \and
Viktor Olejar\inst{4} \and
Cem Nurlu\inst{5} \and
Abuzer Yakary{\i}lmaz\inst{6,7}}
\authorrunning{U. Birkan et al.}
%
\institute{
Department of Computer Engineering, Middle East Technical University, Turkey \and
Department of Physics, Middle East Technical University, Turkey \email{utku.birkan@gmail.com} \and
Department of Computer Science, \"{O}zye\u{g}in University, Turkey \email{ozlemsalehi@gmail.com} \and
Institute of Mathematics, P.J. Šafárik University in Košice, Slovakia \email{viki.olejar@gmail.com} \and
Department of Physics, Boğaziçi University, Turkey
\email{cem.nurlu@gmail.com} \and
Center for Quantum Computer Science, University of Latvia, Latvia
\email{abuzer@lu.lv} \and
QWorld Association, \url{https://qworld.net}
}
\maketitle              
\begin{abstract}
Quantum finite automata (QFAs) literature offers an alternative mathematical model for studying quantum systems with finite memory. As a superiority of quantum computing, QFAs have been shown exponentially more succinct on certain problems such as recognizing the language \(\MODp=\setBuilder{a^{j}}{j \equiv 0 \mod p}\) with bounded error, where $p$ is a prime number. In this paper we present improved circuit based implementations for QFA algorithms recognizing the \MODp\ problem using the Qiskit framework. We focus on the case $p=11$ and provide a 3 qubit implementation for the \MOD{11} problem reducing the total number of required gates using alternative approaches. We run the circuits on real IBM quantum devices but due to the limitation of the real quantum devices in the NISQ era, the results are heavily affected by the noise. This limitation reveals once again the need for algorithms using less amount of resources. Consequently, we consider an alternative 3 qubit implementation which works better in practice and obtain promising results even for the problem \MOD{31}.

\keywords{quantum finite automata \and quantum circuit \and rotation gate \and quantum algorithms.}
\end{abstract}

\section{Introduction}

Quantum finite automata literature offers an alternative mathematical model for studying quantum systems with finite memory. Many different models have been proposed  with varying computational powers \cite{AY15A}. Moore-Crutchfield quantum finite automaton (MCQFA) \cite{MC00} is one of the earliest proposed models which is obtained by replacing the transition matrices of the classical finite automata by unitary operators. Despite the fact that they are weaker than their classical counterparts in terms of their language recognition power, for certain languages MCQFAs have been shown to be more succinct. One such example is the language \(\MODp = \setBuilder{a^{j}}{j \equiv 0 \mod p}\), where $p$ is a prime number: MCQFAs were shown to be exponentially more space-efficient than their classical counterparts \cite{AF98}.

Experimental demonstration of quantum finite automata has recently gained popularity. In \cite{TFL19}, the authors implement an optical quantum finite automaton for solving promise problems. A photonic implementation for \MODp\ problem is presented in \cite{MPC20}. MCQFA for the \MODp\ problem has been also implemented using a circuit based approach within Qiskit and Rigetti frameworks by K\={a}lis in his Master's Thesis \cite{Kalis18}.

As a continuation of \cite{Kalis18}, in this paper we present improved circuit based implementations for MCQFA recognizing the \MODp\ problem using the Qiskit framework. We start with the naive implementation proposed in \cite{Kalis18} and provide a new implementation which reduces both the number of qubits and the number of required basis gates, due to an improved implementation of the multi-controlled rotation gate around $y$-axis and the order in which the gates are applied. We demonstrate the results of the experiments carried out by IBMQ backends for both the improved naive implementation and the optimized implementation of \cite{Kalis18}. Regarding the optimized implementation, we experimentally look for the parameters which would minimize the maximum error rate. 

We also propose a 3-qubit parallel implementation which works better in practice for the \MOD{11} and \MOD{31} problems. The choice of parameters for this implementation heavily influences the outcomes unlike the optimized implementation where this choice does not have a huge impact on the acceptance probabilities. 

We conclude by suggesting a new implementation for the rotation gate around $y$-axis, taking into account the new \emph{basis gates}, the gates that are implemented at the hardware level--that have been recently started to be used by IBM. This new proposal lays the foundations for future work on the subject. The source code of our quantum circuits can be accessed from the link below:
\begin{center}
	\scriptsize 
	\url{https://gitlab.com/qworld/qresearch/research-projects/qfa-implementation/-/tree/iccs-2021}
\end{center}

\section{Background}

We assume that the reader is familiar with the basic concepts and terminology in automata theory and quantum computation. We refer the reader to \cite{NC00,Sip06,AY15A} for details.

Throughout the paper, \(\Sigma\) denotes the finite input alphabet, not containing the left and right-end markers (\textcent\ and \$, respectively), and \(\tilde{\Sigma}\) denotes \(\Sigma \cup \{\cent, \$ \}\). For a string \(w\in \Sigma^{*}\), its length is denoted by \(\abs{w}\) and, if \(\abs{w}>0\), \(w_i\) denotes the \(i\)\textsuperscript{th} symbol of \(w\). For any given input string \(w\), an automaton processes string \(\tilde{w}=\cent w \$\) by reading it symbol by symbol and from left to right.

There are several models of quantum finite automata (QFAs) in the literature with different computational powers \cite{AY15A}. In this paper, we focus on the known most restricted model called as \emph{Moore-Crutchfield quantum finite automaton} (MCQFA) model \cite{MC00}. 

Formally, a \(d\)-state MCQFA is a 5-tuple \[
    M = (\Sigma, Q, \{U_\sigma \mid \sigma \in \tilde{\Sigma}\}, q_s, Q_A),
\] where \(Q=\{q_1,\ldots,q_d\}\) is the finite \emph{set of states}, \(U_\sigma\) is the \emph{unitary operator} for symbol \(\sigma \in \tilde{\Sigma}\), \(q_s \in Q\) is the \emph{start state}, and \(Q_A \subseteq Q\) is the \emph{set of accepting states}.

The computation of \(M\) is traced by a \(d\)-dimensional vector, called the state vector, where \(j\)\textsuperscript{th} entry corresponds to state \(q_j\).  At the beginning of computation, \(M\) is in quantum state \(\ket{q_s}\), a zero vector except its \(s\)\textsuperscript{th} entry, which is 1. For each scanned symbol, say \(\sigma\), \(M\) applies the unitary operator \(U_\sigma\) to the state vector. After reading symbol \$, the state vector is measured in the computational basis. If an accepting state is observed, the input is accepted. Otherwise, the input is rejected.

For a given input \(w \in \Sigma^{*}\), the final state vector \(\ket{v_f}\) is calculated as\[
    \ket{v_f} = U_{\$} U_{w_{|w|}} \cdots U_{w_1} U_{\cent} \ket{q_s}.
\] Let $ \ket{v_f} = \left( \alpha_1 ~~ \alpha_2 ~~ \cdots ~~ \alpha_d \right)^T $. Then, the probability of observing the state \(q_j\) is \(\abs{\alpha_j}^2\), and so, the accepting probability of \(M\) on \(w\) is \(\sum_{q_j \in Q_A} \abs{\alpha_j}^2\).

\section{\MODp\ Problem and QFA Algorithms}

For any prime number $p$, we define language \[
    \MODp = \setBuilder{a^{j}}{j \equiv 0 \mod p}.
\]Ambainis and Freivalds \cite{AF98} showed that MCQFAs are exponentially more succinct than their classical counterparts, i.e., \MODp\ can be recognized by an MCQFA with \(\bigO(\log p)\) states with bounded error, while any probabilistic finite automaton requires at least $p$ states to recognize the same language with bounded error. The MCQFA constructions given in \cite{AF98} were improved later by Ambainis and Nahimovs \cite{AN09}. 

\subsection{2-State QFA}

We start with giving the description of a 2-state MCQFA that accepts each member of \MODp\ with with probability 1 and rejects each nonmember with a nonzero probability.

Let \(M_p\) be an MCQFA with the set of states \(Q=\{ q_1, q_2\}\), where \(q_1\) is the starting state and the only accepting state. The identity operator is applied when reading \textcent\ or \$. Let~$\Sigma = \{a\}$, which is often denoted as a unary alphabet. For each symbol \(a\), the counter-clockwise rotation with angle \({2\pi}/{p}\) on the unit circle is applied:
\[
    U_a =
    \begin{pmatrix}
        \cos{(2\pi/p)} & ~ & -\sin{(2\pi/p)} \\
        \sin{(2\pi/p)} & & \cos{(2\pi/p)} \\
    \end{pmatrix}\text{\,.}
\]

The minimal rejecting probability of a non-member string \(w\) by the automaton \(M_p\) is \(\sin^2 \mypar{{\abs{w}\cdot2\pi}/{p}}\), which gets closer to 0 when \(\abs{w}\) approaches an integer multiple of p. One may notice that instead of \({2 \pi }/{p}\), it's also possible to use the rotation angle \(k\cdot{2 \pi }/{p}\) for some \(k \in \{1,\ldots,p-1\}\). It is easy to see that the rejecting probability of each non-member differs for different values of $k$, but the minimal rejecting probability will not be changed when considering all non-members. 

On the other hand, to obtain a fixed error bound, we can execute more than one 2-state MCQFA in parallel, each of which uses a different rotation angle. 

\subsection{{\(\bigO(\textnormal{\bf log}\, p)\)}-State QFAs}

Here we explain how to combine 2-state MCQFAs with different rotation angles to obtain a fixed error bound.

First, we define the 2-state MCQFA $ M_p^k $ as same as $ M_p $ except for the rotation angle for the symbol $ a $, which is now $ k \cdot {2\pi}/{p} $ where $ k \in \{1,\ldots,p-1\} $.

Then we define the $2d$-state MCQFA $ M_p^K $, where $ K $ is a set formed by $d$ many $k$ values: $ K = \{ k_1,\ldots,k_d \} $ and each $ k_j \in K $ is an integer between 1 and $ p-1 $. The MCQFA $ M_p^K $ executes $ d $ 2-state MCQFAs $ \{ M_p^{k_1},\ldots,M_p^{k_d} \} $ in parallel. The state set of $ M_p^K $ is formed by $ d $ pairs of $ \{q_1,q_2\} $:
\[
    \{ q_1^1,q_2^1,q_1^2,q_2^2,\ldots,q_1^d,q_2^d \}\text{\,.}
\]
The state $ q_1^1 $ is the starting state and the only accepting state. At the beginning of the computation, $ M_p^K $ applies a unitary operator $ U_{\cent} $ when reading the symbol $\cent$ and enters the following superposition:\[
    \ket{q_1^1} \; \xrightarrow{\makebox[8mm]{\(U_{\cent}\)}} \; \frac{1}{\sqrt{d}} \ket{q_1^1} + \frac{1}{\sqrt{d}} \ket{q_1^2}+\cdots + \frac{1}{\sqrt{d}} \ket{q_1^d}\text{\,.}
\] In other words, we can say that \(M_p^K\) enters an equal superposition of 2-state MCQFAs  \(M_p^{k_1},M_p^{k_2},\ldots,M_p^{k_d}\).

Until reading the right end-marker, \(M_p^K\) executes each 2-state sub-automaton, \(M_p^{k_j}\), in parallel, where \(M_p^{k_j}\) rotates with angle \(2 \pi k_j/p\). Thus, the overall unitary matrix of \(M_p^K\) for symbol \(a\) is\[
U_a = \bigoplus_{j=1}^d R_j  = \mymatrix{c c c c}{
        R_1 & 0 & \cdots & 0 \\
        0 & R_2 & \cdots & 0 \\
        \vdots & \vdots &  \ddots & \vdots \\
        0 & 0 & \cdots & R_d
    },
    \label{eq:Ua}
\] where\[
    R_{j} =
    \mymatrix{c@{\hspace{3mm}}c} {
    \cos{\left({2\pi k_j}/{p}\right)} & -\sin{\left({2\pi k_j}/{p}\right)} \\
    \sin{\left({2\pi k_j}/{p}\right)} & \cos{\left({2\pi k_j}/{p}\right)} 
    }.
\]

After reading the symbol \$, we apply the unitary operator \(U_{\$} = U_{\cent}^{-1}\). This overall algorithm gives us an exponential advantage of quantum computation over classical computation for some suitable values for $K$, for each $p$. It was shown \cite{AF98} that, for each $p$, there exists a set of $K$ with \(d=\bigO(\log p)\) elements such that \(M_p^K\) recognizes \MODp\ with a fixed error bound. 

\section{\MODp\ Implementations}

In this section, we present our implementation schema in Qiskit and results on simulators and real machines. 

\subsection{Single Qubit Implementation} 

We start with a single qubit implementation. An example implementation of 2-state MCQFA $ M_7 $ for $\MOD{7}$ is given in Figure \ref{circ:4-1_single} where the input string is $aaa$:

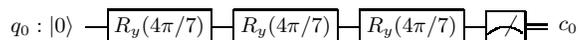
\begin{figure}[h]

\vspace{-0.1in}
    \centering
    \begin{equation*}
        \scl{\(
        \Qcircuit @C=1.0em @R=1em{
            \lstick{q_0:\ket0}  & \gate{R_y(4\pi/7)} & \gate{R_y(4\pi/7)} &  \gate{R_y(4\pi/7)} &  \meter & \rstick{c_0} \cw 
        }\)}
    \end{equation*}
    \caption{Single qubit \MOD7 implementation}
    \label{circ:4-1_single}
\end{figure}

This circuit has one qubit ($q_0$) and one bit ($c_0$). There are different rotation operators (gates) in Qiskit. Here we use $R_y$ gate, which is defined on the Bloch sphere and takes the twice of the rotation angle as its parameter to implement the rotation on the unit circle on the $ \ket{0}-\ket{1} $ plane. The outcome of the measurement at the end is written to the classical bit $c_0$.

\subsection{Three-Qubit Implementations of \MODp\ }

\subsubsection{A Naive Implementation}
To implement the unitary operator given in Equation \eqref{eq:Ua}, we use controlled gates, the conditional statements of the circuits. The implementation cost of the controlled gates are expensive and unfortunately, the straightforward implementation of the above algorithm is costly.

K\={a}lis \cite{Kalis18} gave a four-qubit implementation of the above algorithm for the problem $ \MOD{11} $, where three qubits are used to simulate four sub-automata and one ancilla qubit is used to implement the controlled operators. 

Here we present our implementation schema by using only 3 qubits. All diagrams are obtained by using Qiskit \cite{Qiskit}. We use three qubits called $ q_2, q_1, q_0 $. We implement $ U_{\cent} $ operator by applying Hadamard gates to $ q_2 $ and $ q_1 $. The initial state is $ \ket{000} $. After applying Hadamard operators, we will have the following superposition, in which we represent the state of $q_0$ separately: \[
    \ket{v_{\cent}} = 
    \frac{1}{2} \ket{00} \otimes \ket{0} +
    \frac{1}{2} \ket{01} \otimes \ket{0} +
    \frac{1}{2} \ket{10} \otimes \ket{0} +
    \frac{1}{2} \ket{11} \otimes \ket{0} \\
\] 
The unitary matrix for symbol $a$ is represented as follows:
\[
   U_a =  \mymatrix{cccc}{
    R_1 & 0 & 0 & 0 \\ 
    0 & R_2 & 0 & 0 \\  
    0 & 0 & R_3 & 0 \\ 
    0 & 0 & 0 & R_4 
    }
\]
In order to implement $ U_a $, we apply \(R_1\) when \(q_2 \otimes q_1\) is in state \(\ket{00}\), \(R_2\) when \(q_2 \otimes q_1\) is in state \(\ket{01}\), \(R_3\) when \(q_2 \otimes q_1\) is in state \(\ket{10}\), and \(R_4\) when \(q_2 \otimes q_1\) is in state \(\ket{11}\).

We pick $ K = \{ 1,2,4,8 \} $. Then, after applying $U_a$, the new superposition ($ U_a U_{\cent} \ket{000} $) becomes
\begin{multline*}
    \ket{v_1} =
        \frac{1}{2} \ket{00} \otimes R_y \mypar{\sfrac{2\pi}{11}}\ket{0} +
        \frac{1}{2} \ket{01} \otimes R_y \mypar{\sfrac{4\pi}{11}}\ket{0} + \\
        \dfrac{1}{2} \ket{10} \otimes R_y \mypar{\sfrac{8\pi}{11}}\ket{0} +
        \dfrac{1}{2} \ket{11} \otimes R_y \mypar{\sfrac{16\pi}{11}}\ket{0}.
\end{multline*}
Once we have a block for $ U_a $, then we can repeat it in the circuit as many times as the number of symbols in the input. If our input is $a^m$, then the block for $ U_a $ is repeated $ m $ times. After applying $U_a^m$, the new superposition becomes
\begin{multline*}
    \ket{v_m}=
        \frac{1}{2} \ket{00} \otimes R_y^m \mypar{\sfrac{2\pi}{11}}\ket{0} +
        \frac{1}{2} \ket{01} \otimes R_y^m \mypar{\sfrac{4\pi}{11}}\ket{0} + \\
        \frac{1}{2} \ket{10} \otimes R_y^m \mypar{\sfrac{8\pi}{11}}\ket{0} +
        \frac{1}{2} \ket{11} \otimes R_y^m \mypar{\sfrac{16\pi}{11}}\ket{0}.
\end{multline*}
After reading the whole input, before the measurement, we apply the Hadamard gates which correspond to the operator for symbol \(\$\). 

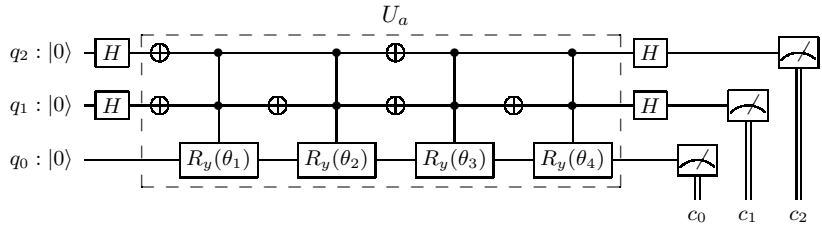
\begin{figure}[h]
\vspace{-0.1in}
    \centering
    \small
    \begin{equation*}
    \scl{\(
    	\Qcircuit @C=.5em @R=1em{
    		&&&&&&&\mbox{\normalsize\(U_a\)}&&&&&&&&\\
    		\lstick{q_2:\ket{0}} & \gate{H} &\qw &\targ & \ctrl{1}              & \qw  & \ctrl{1}              &\targ & \ctrl{1}              & \qw   & \ctrl{1}       &\qw       & \gate{H} & \qw    & \qw    & \meter\\
    		\lstick{q_1:\ket{0}} & \gate{H} &\qw &\targ & \ctrl{1}              &\targ & \ctrl{1}              &\targ & \ctrl{1}              & \targ & \ctrl{1}         &\qw     & \gate{H} & \qw    & \meter \\
    		\lstick{q_0:\ket0}   & \qw   &\qw   & \qw  & \gate{R_y(\theta_1)} &\qw   & \gate{R_y(\theta_2)} &\qw   & \gate{R_y(\theta_3)} & \qw   & \gate{R_y(\theta_4)} & \qw  & \qw    & \meter
    		\gategroup{2}{4}{4}{11}{.8em}{--}  \\
    		&&&&&&&&&&&&&\dstick{c_0}\cwx[-1]&\dstick{c_1}\cwx[-2]&\dstick{c_2}\cwx[-3]
        }\)}
    \end{equation*}
    \caption{A naive \MODp\ circuit implementation}
    \label{circ:modp_naive}
\end{figure}

There is no single-gate solution we can use to implement all of these operators though. Besides, the controlled operators are activated only when all of the control qubits are in state $\ket{1}$. This is why $X$ gates are used; to activate the control qubits when they are in state $\ket{0}$. Figure \ref{circ:modp_naive} depicts a circuit implementing $U_a$.

Initially, two $X$ gates (represented as $\oplus$ in the circuit diagrams) are applied and $ R_1 $ is implemented as the controlled rotation gate will be activated only when the control qubits are initially in state $ \ket{00} $. Next, we apply $X$ gate to $ q_2 $ and so the controlled qubits are activated when in state $ \ket{01} $ and we implement $ R_2 $. Similarly, we implement $ R_3 $ by applying $X$ gates to both qubits so that the controlled qubits are activated when in state $\ket{10}$, and finally we implement $ R_4 $ so that the control qubits are activated in state $\ket{11}$.

\begin{figure}[h]
\vspace{-0.1in}
    \centering
    \small
    \begin{equation*}
    \scl{\(
        \Qcircuit @C=.5em @R=1em{
         &&&&&&\mbox{\normalsize\(U_a\)}&&&&&&&&&&\\
         	\lstick{q_2:\ket0} & \gate{H} & \qw& \ctrl{1} & \qw & \ctrl{1} & \targ & \ctrl{1} & \qw & \ctrl{1} & \targ & \qw & \gate{H}  & \qw & \qw & \qw & \meter \\
         	\lstick{q_1:\ket{0}} & \gate{H} & \qw & \ctrl{1} & \targ & \ctrl{1} & \qw & \ctrl{1} & \targ & \ctrl{1} & \qw & \qw & \gate{H} & \qw & \qw & \meter \\
         	\lstick{q_0:\ket{0}} & \qw & \qw & \gate{R_y(\theta_4)} & \qw & \gate{R_y(\theta_3)} & \qw & \gate{R_y(\theta_1)} & \qw & \gate{R_y(\theta_2)} & \qw 
         	\gategroup{2}{4}{4}{11}{.8em}{--} & \qw & \qw & \qw & \meter \\
         	&&&&&&&&&&&&&&\dstick{c_0}\cwx[-1] &\dstick{c_1}\cwx[-2] & \dstick{c_2}\cwx[-3]\\
         }\)}
    \end{equation*}
    \caption{An improved (but still naive) \MODp\ circuit implementation}
    \label{circ:modp_naive_improved}
\end{figure}
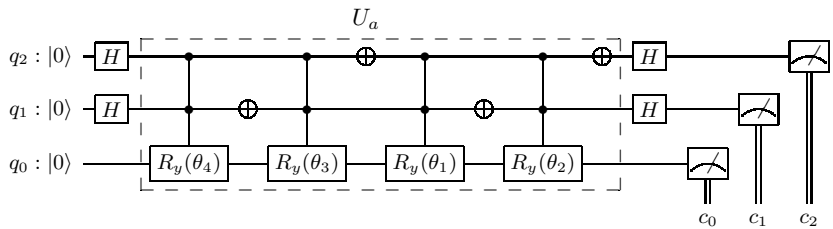

Next, we discuss how to reduce the number of $X$ gates and an improved implementation is given in Figure \ref{circ:modp_naive_improved}. Initially, $ R_4 $ is implemented as the controlled operators will be activated only in state $ \ket{11} $. Next, we apply $X$ gate to $ q_2 $ and so the controlled qubits are activated when in state $ \ket{10} $, and so, we implement $ R_3 $. We apply $X$ gate to $q_1$ and similarly the controlled qubits are activated when in state $ \ket{11} $ so that we implement $ R_1 $. Finally, we apply $X$ gate to $q_2$ again to implement $ R_2 $. Note, that we apply one more $X$ gate at the end so that the initial value of $q_2$ is restored.

The number of $X$ gates can be reduced further by omitting the last $X$ gate in the above diagram and changing the order of $R_y$ gates in the next round so that they follow the values of the controlled qubits. For each scanned $a$, either one of the blocks is applied, alternating between the two, starting with the first block. Overall, three $X$ gates are used instead of four $X$ gates for a single $a$. Note that when the input length is odd, we should always use an extra $X$ gate before the final pair of Hadamard gates. The blocks are depicted below.

\begin{figure}[ht]
\vspace{-0.1in}
    \centering
    \subfloat[First Block]{
        \scl{\(
            \Qcircuit @C=.2em @R=1em{
                & \ctrl{1} & \qw & \ctrl{1} & \targ & \ctrl{1} & \qw & \ctrl{1} & \qw \\
                & \ctrl{1} & \targ & \ctrl{1} & \qw & \ctrl{1} & \targ & \ctrl{1} & \qw\\
                & \gate{R_y(\theta_4)} & \qw & \gate{R_y(\theta_3)} & \qw & \gate{R_y(\theta_1)} & \qw & \gate{R_y(\theta_2)} & \qw
            }\)}
    }
    \hspace{.7cm}
    \subfloat[Second Block]{
        \scl{\(
            \Qcircuit @C=.2em @R=1.1em{
                & \ctrl{1} & \qw & \ctrl{1} & \targ & \ctrl{1} & \qw & \ctrl{1} & \qw \\
                 & \ctrl{1} & \targ & \ctrl{1} & \qw & \ctrl{1} & \targ & \ctrl{1} & \qw\\
                & \gate{R_y(\theta_2)} & \qw & \gate{R_y(\theta_1)} & \qw & \gate{R_y(\theta_3)} & \qw & \gate{R_y(\theta_4)} & \qw 
            }\)}
    }
    \label{circ: modp_naive_blocks}
    \caption{Reducing the NOT gates using two blocks to implement $U_a$}
\end{figure}
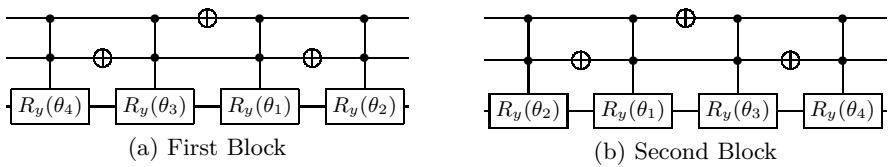

\begin{figure}[h]
\vspace{-0.1in}
    \centering
    \subfloat[K\={a}lis' implementation\label{circ:rygate-kalis}]{
        \adjustbox{margin=1cm 0pt, valign=b}{
        \scl{\(
            \Qcircuit @C=.5em @R=1.1em{
                \lstick{q_2}& \ctrl{1} & \qw      & \ctrl{1} & \qw \\
                \lstick{q_1}& \ctrl{1} & \qw      & \ctrl{1} & \qw \\
                \lstick{\textit{ancilla}}& \targ    & \ctrl{1} & \targ    & \qw     \\
                \lstick{q_0}& \qw & \gate{R_y(\theta)} & \qw & \qw \\
             }\)}
    }}
    \subfloat[Alternative implementation\label{circ:rygate-our}]{
        \adjustbox{margin=1cm 0pt, valign=b}{
        \scl{\(
            \Qcircuit @C=.5em @R=1.1em{
             	 \lstick{q_2}&\qw                 & \ctrl{1}& \qw&  \ctrl{1}&\qw   \\
                \lstick{q_1}&\qw                 & \ctrl{1}& \qw&  \ctrl{1}&\qw     \\
              \lstick{q_0}&\gate{R_y({\theta / 2})}  & \targ & \gate{R_y({-\theta / 2})} &   \targ  & \qw \\
             }\)}}
    }
    \caption{Controlled $R_y$ gate implementations}%
    \label{circ: rygate}
\end{figure}
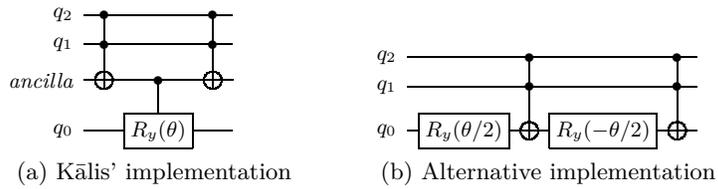

When we implement the circuit in Qiskit, there are different approaches to implement the multi-controlled rotation gate. One option is K\={a}lis' implementation  \cite{Kalis18} that takes advantage of Toffoli gates and controlled rotations as seen in Figure \ref{circ:rygate-kalis}. Another possibility is to use the built-in RYGate class in Qiskit. An alternative implementation of controlled $R_y$ gate presented in \cite{BBC95} is given in Figure \ref{circ:rygate-our}. In this implementation, the rotation gates applied on the target qubit cancel each other unless the control qubits are in state $\ket{11}$ which is checked by the Toffoli gate, thus yielding the same effect as an $R_y$ gate controlled by two qubits. As a further improvement, the Toffoli gate can be replaced with the simplified Toffoli gate, (also referred to as Margolus gate in Qiskit) which has a reduced cost compared to Toffoli gate. This replacement does not affect the overall algorithm as only the states corresponding to first and second sub-automata and that of third and fourth sub-automata are swapped.   

Before running a circuit on an IBMQ device, each gate is decomposed into basis gates $U_1$, $U_2$, $U_3$ and $CX$\footnote{The set of basis gates was changed to $CX$, $I$, $R_z$, $\sqrt{X}$, $X$ in January 2021.} and this decomposition also depends on the backend on which the circuit is run and the physical qubits used on the machine. In the table given below, the number of basis gates required to implement the rotation gate with two controls using the two approaches is given for the IBMQ Santiago and IBMQ Yorktown machines with the default optimization level. 

\begin{table}[h]
    \centering
    \caption{Number of basis gates required by the controlled rotation gate     implementations}
    \begin{tabular}{llrrrrlrrrr}
        \toprule
         & \phantom{ab} & \multicolumn{4}{c}{\texttt{RYGate}} &\phantom{ab}&    \multicolumn{4}{c}{Alternative}\\\cmidrule{3-6}\cmidrule{8-11}
         && $U_1$ & $U_2$ & $U_3$ & $CX$ && $U_1$ & $U_2$ & $U_3$ & $CX$ \\
         \midrule
        Santiago  && 0    & 0    & 6    & 11 && 1    & 2    & 1    & 9 \\
        Yorktown  && 0    & 0    & 6    & 8  && 4    & 1    & 2    & 6 \\
        \bottomrule
    \end{tabular}
\end{table}
Next, we present some experimental results about $\MOD{11}$ problem comparing the performance of the 4 qubit implementation which was originally proposed in \cite{Kalis18} and our improved version with 3 qubits where the number of $X$ gates are reduced and the controlled rotation gates are implemented using the alternative approach. The acceptance probability of each word is the number of times the states $\ket{000}$ and $\ket{0000}$ are observed divided by the number of shots (which was taken as 1000 for the experiments) for 3 qubit and 4 qubit implementations, respectively. The results do not look promising as the acceptance probabilities are around 0.125 and 0.0625 for the 3 qubit and 4 qubit implementations, which are the probabilities of observing a random result. Ideally, the acceptance probabilities for the word lengths 11 and 22 would have been close to 1.

\begin{figure}[htb]
	\centering
	\scalebox{0.9}{\input{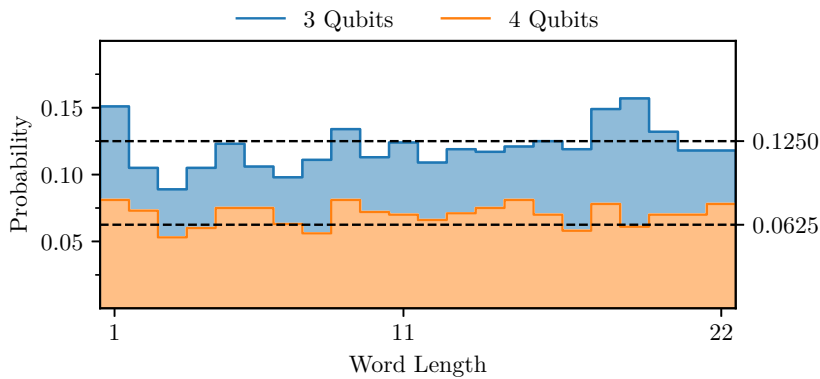}}
	\caption{Acceptance probabilities for \MOD{11} naive implementation}
	\label{graph: 11naive}
\end{figure}

There are three main sources of error in IBMQ machines: number of qubits, circuit depth and the number of \(CX\) gates. Even though our 3-qubits naive implementation reduced the number of qubits and \(CX\) gates, this improvement was not enough to have any meaningful result. Each Margolous gate still requires 3 \(CX\) gates which is better compared to the Toffoli gate which requires 6 but it is still not enough\cite{Qiskit}. In addition, the connectivity of the underlying hardware requires some additional $CX$ gates when the 4-qubits circuit is transpiled. Nevertheless, our implementation provides a significant improvement in the number of basis gates required for the implementation of the algorithm. In Table \ref{tab:gates}, we list the number of basis gates required by both implementations for word length 11 using the default optimization level by IBMQ Santiago backend.

\subsubsection{An Optimized Implementation}

\begin{figure}[ht]
\vspace{-0.1in}
    \small
    \centering
    \scl{\input{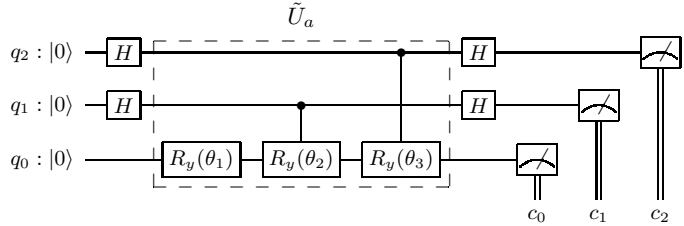}}
    \caption{Optimized implementation for \MOD{11}}
    \label{circ:modp_optimized}
\end{figure}

The circuit construction above can be improved by sacrificing some freedom in the selection of rotation angles as proposed in \cite{Kalis18}. In the circuit diagram given in Figure \ref{circ:modp_optimized}, only the controlled rotation operators are used, where the unitary matrix for symbol $a$ is as follows:
\[
    \tilde{U}_a = \mymatrix{cccc}{
        R_1 & 0 & 0 & 0 \\ 
        0 &  R_2 R_1 & 0 & 0 \\
        0 & 0 & R_3 R_1 & 0 \\
        0 & 0 & 0 & R_3 R_2 R_1
    }
\]
Thus, each sub-automaton applies a combination of rotations among three rotations.

Upon reading the left end-marker, we have the superposition state $\ket{v_{\cent}}$. The block $\tilde{U}_a$ between the Hadamard operators represent the unitary operator corresponding to symbol $a$. After reading the first $a$, the new superposition becomes
\begin{multline*}
    \ket{\tilde{v}_1} = 
        \frac{1}{2} \ket{00} \otimes R_y(\theta_1)\ket{0} +
        \frac{1}{2} \ket{01} \otimes R_y(\theta_2)R_y(\theta_1)\ket{0} +\\
        \frac{1}{2} \ket{10} \otimes R_y(\theta_3)R_y(\theta_1)\ket{0} +
        \frac{1}{2} \ket{11} \otimes R_y(\theta_3)R_y(\theta_2)R_y(\theta_1)\ket{0}.
\end{multline*}

By letting \(\phi_1 = \theta_1\), \(\phi_2 = \theta_1+\theta_2\), \(\phi_3 = \theta_1+\theta_3\), and \(\phi_4 = \theta_1 +\theta_2+\theta_3\), the state \(\ket{\tilde{v}_1}\) can be equivalently expressed as 
\begin{multline*}
    \ket{\tilde{v}_1} = 
    \frac{1}{2} \ket{00} \otimes R_y(\phi_1)\ket{0} + 
    \frac{1}{2} \ket{01} \otimes R_y(\phi_2)\ket{0} + \\ 
    \frac{1}{2} \ket{10} \otimes R_y(\phi_3)\ket{0} +
    \frac{1}{2} \ket{11} \otimes R_y(\phi_4)\ket{0}.
\end{multline*}

Compared to the 3 qubit implementation presented in the previous subsection, this implementation uses less number of gates and especially the number of $CX$ gates is reduced. Furthermore, as no multi-controlled gates are required, the number of $CX$ gates used by the IBMQ Santiago machine even reduces when the default optimization is used. The number of required basis gates for various implementations is given in Table \ref{tab:gates} for word length 11.

\begin{table}[h]
    \centering
    \caption{Number of basis gates required by naive and optimized implementations ran on IBMQ Santiago\label{tab:gates}}
    \begin{tabular}{l@{\hspace{5mm}}rrrr}
    \toprule
         & \multicolumn{4}{c}{Basis Gates} \\\cmidrule{2-5}
        Implementation & $U_1$ & $U_2$ & $U_3$ & $CX$ \\
        \midrule
        3 Qubits Naive & 270 & 15  & 121 & 270  \\[.8ex]
        4 Qubits Naive & 561 & 154 & 114 & 1471 \\[.8ex]
        Optimized      & 0   & 4   & 44  & 55   \\
        \bottomrule
    \end{tabular}
\end{table}

\begin{figure}[htb]
	\centering
    \scl{\input{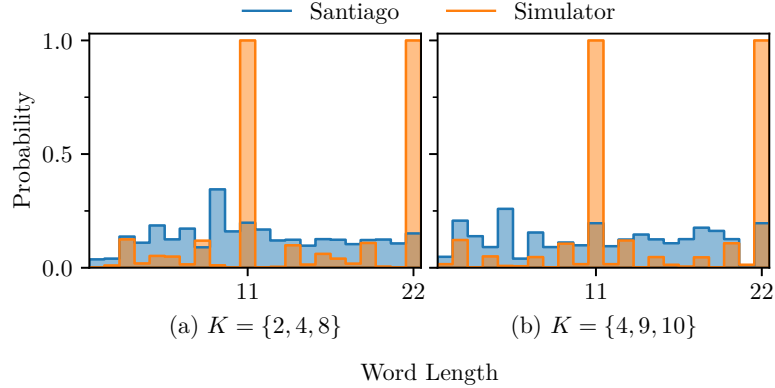}}
	\caption{Acceptance probabilities for \MOD{11} optimized implementation}
	\label{graph: 11optimized}
\end{figure}

We conducted experiments on IBMQ Santiago machine with two different set of values of $k$, $\{2,4,8\}$ and $\{4,9,10\}$. A discussion about the choice of the value of $k$ is presented in the next subsection. The results are still far from the ideal as it can be seen in Figure \ref{graph: 11optimized}. We also plotted the ideal results we got from the simulator. When compared with the naive implementation, we observe that the acceptance probabilities fluctuate until a certain word length but after some point they tend to converge to \sfrac{1}{8}, the probability of selecting a random state.

\subsubsection{A Parallel Implementation}
Recall that in the single qubit implementation, we have a single automaton, but the problem is, that we get arbitrarily large error for nonmember strings and we try to reduce it by running multiple sub-automata in parallel. In this section we provide some experimental results about \MODp\ problem where we simply run each sub-automaton using a single qubit. Although theoretically this approach has no memory advantage, it works better in real devices as no controlled gates are used while still providing a space advantage in terms of number of bits in practice.

\begin{figure}[htb]
\vspace{-0.1in}
    \centering
    \scl{\input{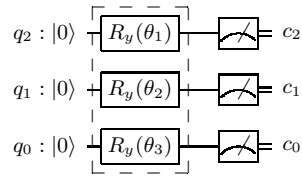}}
    \caption{Parallel \MOD{11} circuit where each qubit implements an MCQFA}
    \label{circ:modp_parallel}
\end{figure}

Consider the circuit diagram given in Figure \ref{circ:modp_parallel}. Using three qubits, we run three automata in parallel with three different rotation angles.

In this implementation, the unitary operators corresponding to \textcent\ and \$\ are identity operators. Upon reading the first $a$, the new state becomes \[
    \ket{v'_1}=R_y(\theta_1)\ket{0}\otimes R_y(\theta_2)\ket{0} \otimes R_y(\theta_3)\ket{0}.
\]

\begin{figure}[htb]
    \centering
    \scalebox{0.9}{\input{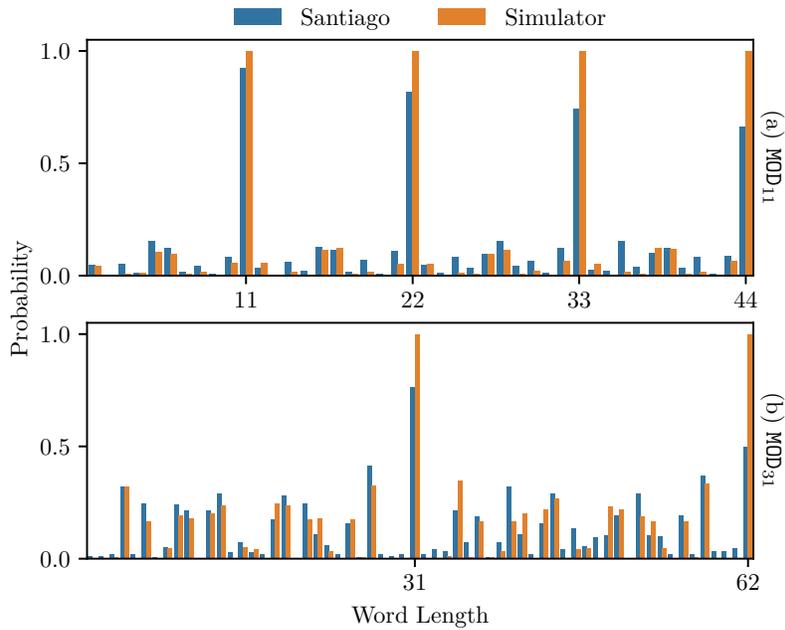}}
    \caption{Acceptance probabilities for \MOD{31} and \MOD{11} parallel implementations}
    \label{graph:parallel}
\end{figure}

We conducted experiments on IBMQ Santiago machine for \MOD{11} problem with $K=\{1,2,4\}$ and for \MOD{31} problem with $K=\{8,12,26 \}$. The results are summarized in Figure \ref{graph:parallel}.

From the graphs above we can see that the experimental results on real machines coincide with the simulator outcome especially for small word lengths. The number of required basis gates to implement the parallel implementation is simply three times the length of the word. 

\subsubsection{Choosing Values of $k$}

In \cite{AN09}, the authors consider various values of $k$ for the optimized implementation. One proposal is the cyclic sequences which work well in numerical experiments and give an MCQFA with $\bigO(\log p)$ states. Another proposal is the AIKPS sequences \cite{AIKPS90} for which the authors provide a rigorous proof but it requires larger number of states. 

We conducted several experiments on the local simulator to see which values of $k$ produce better results for the optimized and parallel MCQFA implementations. We define the maximum error as the highest acceptance probability for a nonmember string and we investigated for which values of $k$, the maximum error is minimized. Experimental findings are listed below, in Table \ref{tab:k_comparison}.

\begin{table}[h]
    \centering
    \caption{Standard deviation and mean values for compared circuits}
    \label{tab:k_comparison}
    \begin{tabular}{l@{\hspace{4mm}}l l l l}
        \toprule
         &  Min & Max & Mean & Std. Dev. \\
        \midrule
        \MOD{11} Optimized & 0.010 & 0.080 & 0.034 & 0.018 \\
        \MOD{11} Parallel  & 0.109 & 0.611 & 0.270 & 0.148 \\
        \MOD{31} Parallel  & 0.319 & 0.974 & 0.615 & 0.167 \\
        \bottomrule
    \end{tabular}
\end{table}

In the optimized implementation, the maximum error ranges between 0.01 and 0.08 with a standard deviation of 0.018 so it can be concluded that the choice of different values of $k$ does not have a significant impact on the success probabilities, for this specific case. When we move on to parallel implementation, we observe that the maximum error probabilities vary heavily depending on the choice of the value of $k$. Interquartile range is much larger this time. Furthermore, \MOD{31} results include error values as high as 0.97 in the third quartile. \MOD{11} also has some outliers that show far greater error than we would like to work with. With this insight, we picked the values of $k$ accordingly in the parallel implementation, which had an impact on the quality of the results we obtained.

\section{Conclusion and Future Work}
The goal of this study was to investigate circuit implementations for quantum automata algorithms solving \MODp\ problem. As a way of dealing with the limitations of NISQ devices, we considered different implementation ideas that reduce the number of gates and qubits used. Our findings contribute to the growing field of research on efficient implementations of quantum algorithms using limited memory.

\begin{figure}[ht]
\vspace{-0.1in}
    \centering
    \subfloat[Qiskit transpilation\label{circ:ry_qiskit}]{
        \adjustbox{margin=1cm 0pt, valign=b}{
        \scl{\(
            \Qcircuit @C=.6em {
                & \gate{\sqrt{X}} & \gate{R_z(\theta+\pi)} & \gate{\sqrt{X}} & \gate{R_z(\pi)} & \qw &
                  \gate{\sqrt{X}} & \gate{R_z(\theta+\pi)} & \gate{\sqrt{X}} & \gate{R_z(\pi)} & \qw
                \gategroup{1}{2}{1}{5}{.8em}{--}
                \gategroup{1}{7}{1}{10}{.8em}{--}
             }\)}
    }}
    
    \subfloat[Alternative implementation\label{circ:ry_new}]{
        \adjustbox{margin=1cm 0pt, valign=b}{
        \scl{\(
            \Qcircuit @C=.6em{
                & \gate{\sqrt{X}} & \gate{R_z(\theta)} & \gate{R_z(\theta)} & \gate{\sqrt{X}} & \gate{X} & \qw
             }\)}}
    }
    \caption{$R_y$ gate implementations with the new basis set}%
\end{figure}
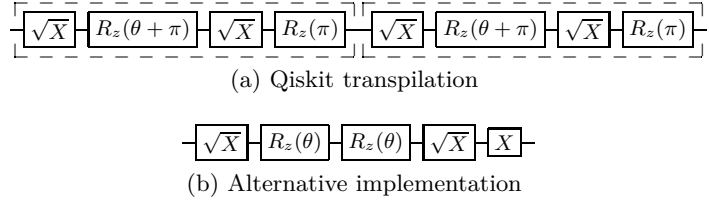

Recently, the basis gates of IBMQ backends were reconfigured as $CX$, $I$, $R_z$, $\sqrt{X}$, and $X$. As a result, a new methodology should be developed in order to reduce the number of required gates. For instance, the circuit with two consecutive $R_y$ gates that is transpiled by Qiskit is given in Figure \ref{circ:ry_qiskit}. Instead of this design, we propose the implementation given in Figure \ref{circ:ry_new}, which would reduce the number of required basis gates. We use the fact that \(R_y(\theta) = \sqrt{X} \cdot R_z(\theta) \cdot \sqrt{X}\cdot X\), hence multiple rotations can be expressed as \(R^n_y(\theta) =  \sqrt{X} \cdot R^n_z(\theta) \cdot \sqrt{X}\cdot X\).

\section*{Acknowledgements}
This research project started during QIntern2020 program of QWorld (July-August 2020) and later continued under the QResearch Department of QWorld.

We thank to anonymous reviewers for their helpful comments and M\={a}rti\c{n}\v{s} K\={a}lis for giving a presentation during QIntern2020 program on the details of his master thesis and kindly answering our questions. 

Birkan and Nurlu were partially supported by T\"{U}B\.{I}TAK scholarship ``2205 -- Undergraduate Scholarship Program''. Yakary{\i}lmaz was partially supported by the ERDF project Nr. 1.1.1.5/19/A/005 ``Quantum computers with constant memory''.

\bibliographystyle{splncs04}
\bibliography{lit}

\end{document}